\documentclass[twocolumn,aps,prl,floatfix,superscriptaddress]{revtex4-1}

\usepackage{amsmath,amssymb}
\usepackage{graphicx,float}
\usepackage{times,txfonts}
\usepackage{color}
\usepackage{multirow}
\usepackage{bbold}
\usepackage{color}
\usepackage{xcolor}
\usepackage{soul}
\usepackage{hyperref}
\usepackage{times,txfonts}
\usepackage{natbib}
\usepackage{nicefrac}
\usepackage{blindtext}
\usepackage[export]{adjustbox}
\usepackage{mathrsfs}
\usepackage{textcomp}
\usepackage{multirow}
\usepackage[normalem]{ulem}
\usepackage{amsmath}

    \newcommand{\ii}{\text{i}}
\newcommand{\dd}{\mathrm{d}}
\newcommand{\ee}{\mathrm{e}}

\renewcommand{\epsilon}{\varepsilon}

\def\VR{\kern-\arraycolsep\strut\vrule &\kern-\arraycolsep}
\def\vr{\kern-\arraycolsep & \kern-\arraycolsep}
\usepackage{soul}
\definecolor{lightblue}{RGB}{185,210,248}
\definecolor{lgreen}{RGB}{15,150,15}
\sethlcolor{lightblue}

\begin{document}

\title{Optimal Diffractive Focusing of Vortex Beams}

\author{Roohollah Ghobadi}
\email{farid.ghobadi80@gmail.com}
\affiliation{Institute for Quantum Studies, Chapman University, Orange, California 92866, USA}
\affiliation{Nexus for Quantum Technologies, University of Ottawa, Ottawa, ON, Canada, K1N 5N6}

\author{Carrie Rodriguez}
\affiliation{Institute for Quantum Studies, Chapman University, Orange, California 92866, USA}

\author{Lukas Scarfe}
\affiliation{Nexus for Quantum Technologies, University of Ottawa, Ottawa, ON, Canada, K1N 5N6}

\author{Maxim A. Efremov}
\affiliation{German Aerospace Center (DLR), Institute of Quantum Technologies, 89081 Ulm, Germany}

\author{Ebrahim~Karimi}
\affiliation{Institute for Quantum Studies, Chapman University, Orange, California 92866, USA}
\affiliation{Nexus for Quantum Technologies, University of Ottawa, Ottawa, ON, Canada, K1N 5N6}

\date{\today}

\begin{abstract}
We formulate diffractive ring focusing of paraxial beams carrying orbital angular momentum (OAM) through a finite circular aperture as a variational optimization problem. For a fixed topological charge $\ell$, we determine the real-valued radial pupil that maximizes the intensity at a prescribed ring radius and propagation distance under a normalization constraint on the input field. The resulting problem admits an exact analytical solution: the optimal pupil combines the familiar Fresnel phase structure with an OAM-dependent Bessel radial modulation, and the associated eigenvalue yields a rigorous upper bound on the achievable ring intensity. We further analyze the lens-assisted configuration and derive a simple bound that allows direct comparison between diffractive and refractive vortex focusing. The theory also provides a quantitative framework for evaluating the performance of practical mask designs, including binary Fresnel–Bessel zone plates and amplitude-weighted pupils. We have experimentally tested our predictions in the optical domain by implementing the corresponding vortex-beam pupils and comparing their propagation and focusing performance with the optimal solution. Our results establish the vortex-beam analogue of optimal amplitude-only focusing and identify the fundamental limits governing diffractive concentration of structured waves.
\end{abstract}

\maketitle

\noindent\textit{Introduction---} The ability to concentrate wave energy at a prescribed location, i.e. focusing, lies at the heart of many areas of science and technology. In optics, acoustics, and matter-wave physics, focusing is typically achieved using lenses or mirrors that shape the phase or arrival time of a propagating field so that different portions of the wavefront arrive in phase and at the same time at the desired focal point~\cite{born2013principles}. Achieving such control requires adjusting optical paths with sub-wavelength precision, which becomes increasingly challenging as the wavelength decreases. In regimes ranging from X-rays to sub-picometer matter waves, suitable materials may interact only weakly with the field, while fabrication tolerances approach fundamental limits~\cite{Attwood1999}. Broadband or ultrashort pulses introduce an additional complication due to dispersion, which can distort the beam's temporal structure during propagation~\cite{karimi2013influence}. 

Diffractive approaches offer an alternative to conventional refractive focusing that does not rely on refractive elements. Fresnel's zone plate, consisting of alternating transparent and opaque rings, demonstrated early on that constructive interference of diffracted waves can generate a bright focal spot even when the field is shaped only through spatial amplitude modulation~\cite{myers1951studies,sussman1960elementary}. Since then, diffractive optical elements such as kinoforms, holographic masks and amplitude apertures have become powerful tools for controlling wave propagation~\cite{Brown:66,Lesem1969}. Importantly, these concepts extend beyond light to other wave systems, including electrons and neutrons, where conventional lenses/phase elements are difficult to realize~\cite{verbeeck2010production,clark2015controlling,larocque2018twisted}. In such contexts, shaping the field within a finite aperture can provide a practical lens-free method for concentrating wave intensity~\cite{cirone2001quantum,bialynicki2002and,case2012diffractive,weisman2017diffractive,weisman2021diffractive}. Despite their widespread use, the fundamental limits of diffractive focusing have only recently been examined from an optimization perspective. Efremov et al.~\cite{efremov2025optimal} posed a benchmark problem: among all real-valued, aperture-limited and normalized input fields, determine the pupil function that maximizes the intensity at a prescribed propagation distance. Their variational treatment yields an optimal pupil profile together with a rigorous upper bound on the achievable focal intensity. Remarkably, the analysis shows that conventional Fresnel-zone constructions are not optimal and that properly designed amplitude modulation can enhance the attainable focal intensity by a factor of $(\pi/2)^2$ relative to standard amplitude Fresnel zone plates. These results establish a principled reference point for evaluating diffractive focusing strategies. \newline
A distinct situation arises for waves carrying orbital angular momentum (OAM). Beams with a helical phase structure $\ee^{\ii\ell\varphi}$, $\ell$ is the topological charge and an integer number, and $\varphi$ is the azimuth angle of polar coordinates, possess a phase singularity on the optical axis, forcing the intensity to vanish at the center for any non-zero topological charge $\ell$~\cite{rubinsztein2017roadmap}. Consequently, the natural focusing objective is no longer the formation of a bright point but rather the concentration of energy onto an annular ring whose radius depends on both the OAM charge and the radial field profile. Vortex beams have attracted considerable attention in recent decades because of their unique interaction with matter and their growing role in applications such as optical trapping, high-resolution microscopy, nonlinear optics and strong-field physics. In particular, maximizing the intensity along the characteristic doughnut-shaped ring while preserving the central phase singularity is essential for techniques such as stimulated emission depletion microscopy~\cite{Hell:94}, for nonlinear processes including multiphoton absorption~\cite{sederberg2020vectorized,begin2023nonlinear}, probing materials' chirality/dichroism, filamentation~\cite{bouchard2016polarization}, and high-harmonic generation~\cite{kong2017controlling}. 

Here, we investigate the diffractive focusing of aperture-limited vortex beams. We formulate the problem as an optimization over \textit{real} radial pupil functions that shape an incoming OAM mode within a finite aperture, with the objective of maximizing the intensity at a prescribed ring radius and propagation distance. This choice reflects the natural focusing geometry imposed by the phase singularity of vortex beams. The resulting formulation leads to an integral-kernel eigenvalue problem, with the principal eigenfunction specifying the optimal pupil profile and the corresponding eigenvalue determining the maximum achievable ring intensity. Beyond providing a practical design rule, this framework establishes a fundamental performance limit for diffractive focusing of aperture-limited vortex beams. In particular, it identifies the pupil structures that maximize energy concentration on the characteristic annular intensity distribution and clarifies the role of radial field shaping in vortex-beam propagation. The results, therefore, provide the vortex-beam analogue of the optimal amplitude-only focusing problem previously established for conventional beams. By doing so, they offer a rigorous benchmark for evaluating diffractive vortex optics and enable direct comparisons with conventional vortex focusing techniques based on refractive elements, spiral phase plates and zone-plate structures. More broadly, our results reveal how the interplay between Fresnel diffraction and the radial structure imposed by OAM determines the ultimate limits of intensity concentration in vortex beams. This perspective provides a unified framework for analyzing and designing diffractive focusing schemes for structured waves, with potential implications for applications ranging from nonlinear optics and strong-field physics to high-resolution microscopy and structured matter-wave beams.\\

\noindent\textit{Theory ---} We consider an aperture-limited field carrying a definite OAM value $\ell$, whose transverse profile at the input plane can be written as
\begin{equation}
	\Psi_0(\rho,\varphi)=\psi_\ell(\rho)\ee^{\ii\ell\varphi}.
\end{equation}
Here $\rho=r/R$ denotes the dimensionless radial coordinate, $0\leq \rho\leq 1$, within a circular aperture of the radius $R$, and $\psi_\ell(\rho)$ is a real radial pupil function that specifies the amplitude distribution across the aperture. In general, the radial profile $\psi_\ell(\rho)$ depends on the magnitude of the OAM index, $|\ell|$, and is identical for opposite signs of $\ell$, i.e., $\psi_\ell(\rho)=\psi_{|\ell|}(\rho)$. The propagation of such a field in a paraxial regime, where the spatial extension $w$ is much greater than the wavelength, i.e. $w\gg\lambda$, is given by the paraxial wave (Helmholtz) equation,
\begin{equation}
	\ii\frac{\partial}{\partial \tau}\Psi(\rho,\varphi, \tau)=-\frac{1}{2}\nabla^2\Psi(\rho,\varphi, \tau),
\end{equation}
which has the same form of a free-particle scalar Schr\"odinger equation. Here, $\tau$ is equivalent to the longitudinal distance $z=k R^2\tau$ from the aperture, where $k$ denotes the wave number, and $\nabla^2$ stands for a 2D Laplacian. Rotational symmetry preserves the OAM sector during free-space propagation, so that a single $\ell$ field remains of the form $\Psi(\rho,\varphi,\tau)=\ee^{i\ell\varphi}\psi(\rho,\tau)$, with the radial component obeying 
\begin{equation} \label{psiprop}
	\Psi(\rho,\tau)=\frac{(-\ii)^{\ell+1}}{\tau}\ee^{\ii\frac{\rho^2}{2\tau}} \int_0^1 \rho'\,\dd\rho'\,\ee^{\ii\frac{\rho'^2}{2\tau}}\,J_\ell\left(\frac{\rho\rho'}{\tau}\right)\,\psi_\ell(\rho'),
\end{equation}
where $J_\ell(x)$ is the Bessel function of the first kind of the order $\ell$ (see Appendix for more details). Our goal is to determine the pupil function $\psi_\ell(\rho)$ such that maximizes the intensity at a prescribed ring radius $\rho^*$ at a given focusing time $\tau_{\rm f}$. Physically, this corresponds to concentrating the energy of the vortex beam onto an annulus, which is the natural focusing geometry imposed by the phase singularity of OAM beams.  We, therefore, seek to maximize the functional $J[\psi_\ell]=|\Psi(\rho^*,\tau_{\rm f})|^2$, subject to the normalization constraint on the input field. Introducing a Lagrange multiplier $\Lambda$ to enforce this constraint leads to the variational functional
\begin{equation}\label{var}
	\mathcal L[\psi_\ell]=J[\psi_\ell]-\Lambda\left[2\pi\int_0^1 \rho \dd\rho\, \psi_\ell^2(\rho)-1\right].
\end{equation}
Using Eq.~\eqref{psiprop}, the objective functional can be expressed as a quadratic form in the pupil function,
\begin{equation}\label{int}
	J[\psi_\ell]=\int_0^1\rho\, \dd\rho\int_0^1\rho'\,\dd\rho'\,\mathcal K_{\ell}(\rho,\rho')\,\psi_\ell(\rho)\,\psi_\ell(\rho'),
\end{equation}
where the real symmetric kernel
\begin{equation}\label{kernel}
	\mathcal K_{\ell}(\rho,\rho')=\frac{1}{\tau_{\rm f}^2}\cos\!\left(\frac{\rho^2-\rho'^2}{2\tau_{\rm f}}\right)J_\ell\!\left(\frac{\rho^*\rho}{\tau_{\rm f}}\right)\,J_\ell\!\left(\frac{\rho^* \rho'}{\tau_{\rm f}}\right),
\end{equation}
encodes the propagation dynamics and the target focusing radius $\rho^*$. For $\ell=0$ and on-axis focusing ($\rho^*=0$), Eq.~\eqref{kernel} reduces to $\mathcal K_{0}(\rho,\rho')=(1/\tau_{\rm f}^2)\cos\!\left[\left(\rho^2-\rho'^2\right)/(2\tau_{\rm f})\right],$ thus recovering the kernel of Ref.~\cite{efremov2025optimal}. Variation of Eq.~\eqref{var} with respect to variations of $\psi_\ell(\rho)$ yields eigenvalue problem
\begin{equation}
	\frac{1}{2\pi} \int_0^1\rho'\,\dd\rho'\,\mathcal K_{\ell}(\rho,\rho')\,\psi_\ell(\rho')=\Lambda\,\psi_\ell(\rho),
\label{eigenvalue}
\end{equation}
where the largest eigenvalue $\Lambda$ determines the maximum achievable ring intensity. As detailed in Appendix, the eigenvalues of Eq.~(\ref{eigenvalue}) are given by

\begin{equation}\label{jmax}
	\Lambda_{\pm}^{(\ell)}(\rho^*,\tau_{\mathrm f})=\frac{1}{2\tau_{\rm f}^2}\left(\mathcal I_{\ell} \pm |\mathcal R_{\ell}|\right),
\end{equation}
with 
\begin{align}\label{IZ}
	\mathcal I_{\ell}(\rho^*,\tau_{\mathrm f})=&\frac{1}{2\pi}\int_0^1 \rho\,\dd\rho\,J_\ell^2\left(\frac{\rho^*\rho}{\tau_{\rm f}}\right),\\ \label{R}
	\mathcal R_{\ell}(\rho^*,\tau_{\mathrm f})=&\frac{1}{2\pi}\int_0^1\rho\,\dd\rho\,J_\ell^2\left(\frac{\rho^*\rho}{\tau_{\rm f}}\right)\,e^{\ii\frac{\rho^2}{\tau_{\rm f}}}.
\end{align}
Here $\Lambda^{(\ell)}_+=\Big(\mathcal I_{\ell} +|\mathcal R_{\ell}|\Big)/{(2\tau_{\rm f}^2)}$ corresponds to the maximum achievable ring intensity. The corresponding normalized eigenfunctions read
\begin{align}\label{psipml}
	\psi_\ell^{+}(\rho)&=\frac{1}{\sqrt{\tau_{\rm f}^2\,\Lambda^{(\ell)}_+}}\,J_\ell\left(\frac{\rho^*\rho}{\tau_{\rm f}}\right)\cos\left(\frac{\rho^2}{2\tau_{\rm f}}-\theta_{\ell}\right),\\ \nonumber
	\psi_\ell^{-}(\rho)&=\frac{1}{\sqrt{\tau_{\rm f}^2\,\Lambda^{(\ell)}_-}}J_\ell\left(\frac{\rho^*\rho}{\tau_{\rm f}}\right)\sin\left(\frac{\rho^2}{2\tau_{\rm f}}-\theta_{\ell}\right),
\end{align}
with the phase shift $\theta_{\ell}=(1/2)\arg(\mathcal R_{\ell})$. Equations~\eqref{jmax}-\eqref{psipml} are the main results of this work, and in the following, we discuss their consequences. 

Inspecting Eqs.~\eqref{jmax}-\eqref{psipml} shows that $\Lambda^{(\ell)}_\pm$ is invariant under $\ell\to -\ell$ transformation. Accordingly, $\psi_\ell^{+}(\rho) \ee^{\pm\ii\ell\varphi}$ and their superposition, {\it e.g.} $\psi_\ell^{+}(\rho)\cos(\ell\varphi)$, have the same total focusing merit, but the latter one concentrates intensity into $2|\ell|$ angular lobes. This opens potential applications such as the trapping of atoms in an optimally designed ring lattice or the symmetry-sensitive probing of perturbations.

\begin{figure}
    \centering
    \includegraphics[width=\linewidth]{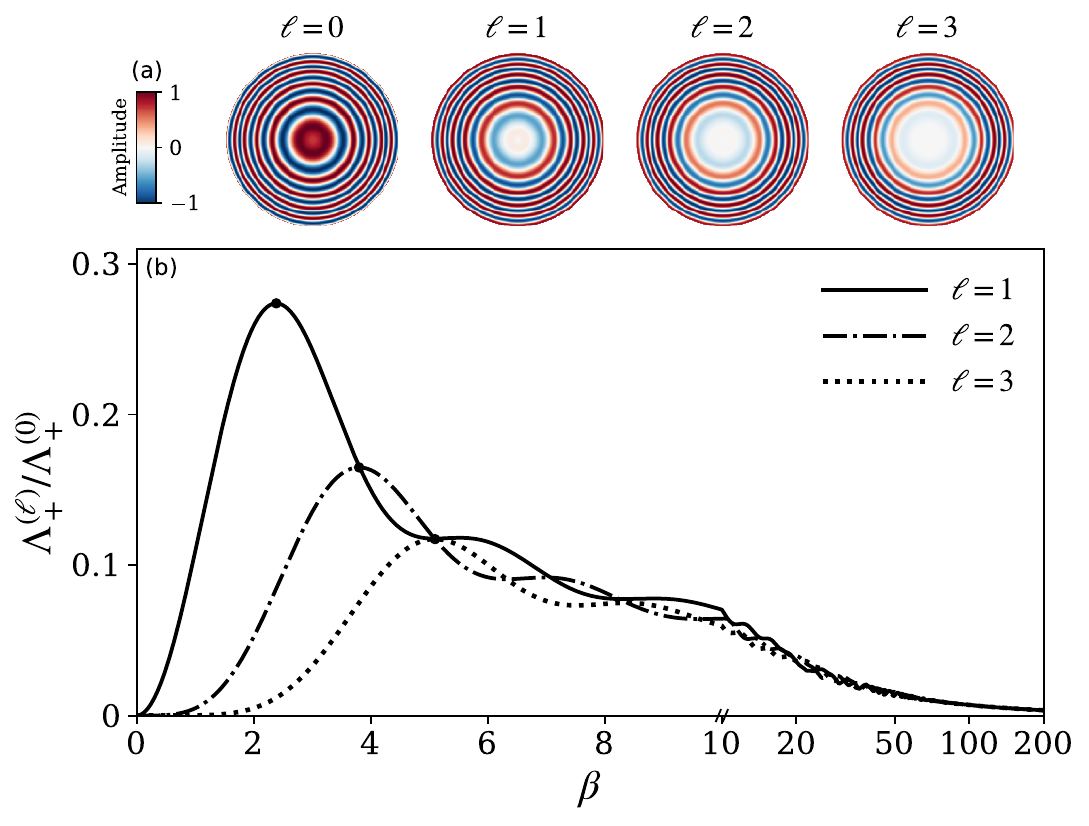}
    \caption{OAM dependence of the optimal focusing solution. (a) Optimal normalized eigenfunction $\psi^{+}_{\ell}(\rho)$ for $\ell=0,1,2,3$. (b) Relative focusing strength
    $\Lambda^{(\ell)}_{+}/\Lambda^{(0)}_{+}$ versus the dimensionless radial parameter $\beta={\rho^{*}}/{\tau_{\rm f}}$, for $\ell=1$ (solid), $\ell=1$ (dashed-dotted) and $\ell=3$ (dotted). The horizontal axis is linear for $0\leq\beta\leq 10$ and logarithmically compressed for $\beta>10$ to display the asymptotic regime. The peak shifts to larger $\beta$ with increasing $\ell$, reflecting the outward displacement of the dominant Bessel weight $J_\ell^2(\beta\rho)$ for higher-order OAM modes.
    }
    \label{fig:lambdap}
\end{figure}

Figure~\ref{fig:lambdap} shows how OAM modifies the optimal focusing mode and its efficiency. We consider the Fresnel times $\tau_n=(2\pi n)^{-1}$, for which the quadratic propagation phase spans $n$ Fresnel zones across the pupil. Panel (a) shows the optimal pupil amplitudes for $\ell=0,1,2,3$, where increasing $\ell$ shifts the radial weight outward the Bessel factor $J_{\ell}$. Panel (b) shows the corresponding relative focusing strength $\Lambda^{(\ell)}_{+}/\Lambda^{(0)}_{+}$. In the fast Fresnel limit, $\tau_n\ll1$, where $\mathcal{R}_{\ell}\approx0$, the maximum occurs near $\beta_{\text{max}}(\ell)\simeq j_{l-1,1}$, with $j_{\nu,1}$ the first zero of $J_\nu$. At large $\beta$, relative focusing strength approach the common $\beta^{-1}$ decay.\\

\noindent\textit{Focusing performance of arbitrary pupil functions.} Next, we quantify the focusing performance of an arbitrary normalised pupil relative to the optimal bound. To this end, we invoke the spectral decomposition to write the integral kernel in Eq.~\eqref{kernel} as
\begin{equation}
	\mathcal K_\ell=\Lambda^{(\ell)}_{+}\,|\psi_\ell^{+}\rangle\langle\psi_\ell^{+}|+\Lambda^{(\ell)}_{-}\,|\psi_\ell^{-}\rangle\langle\psi_\ell^{-}|,
\end{equation}
where $\psi_\ell^{\pm}$ are the normalised eigenfunctions obtained in Eq.~\eqref{psipml}. For an arbitrary normalised pupil function $\psi$, we define focusing efficiency 
\begin{equation}\label{etadef}
	\eta_{\ell}(\psi)=\frac{J[\psi]}{\Lambda^{(\ell)}_{+}}=\frac{\langle\psi,\mathcal{K}_\ell\psi\rangle}{\Lambda^{(\ell)}_{+}}.
\end{equation}

\begin{figure*}
    \centering
    \includegraphics[width=0.9\textwidth]{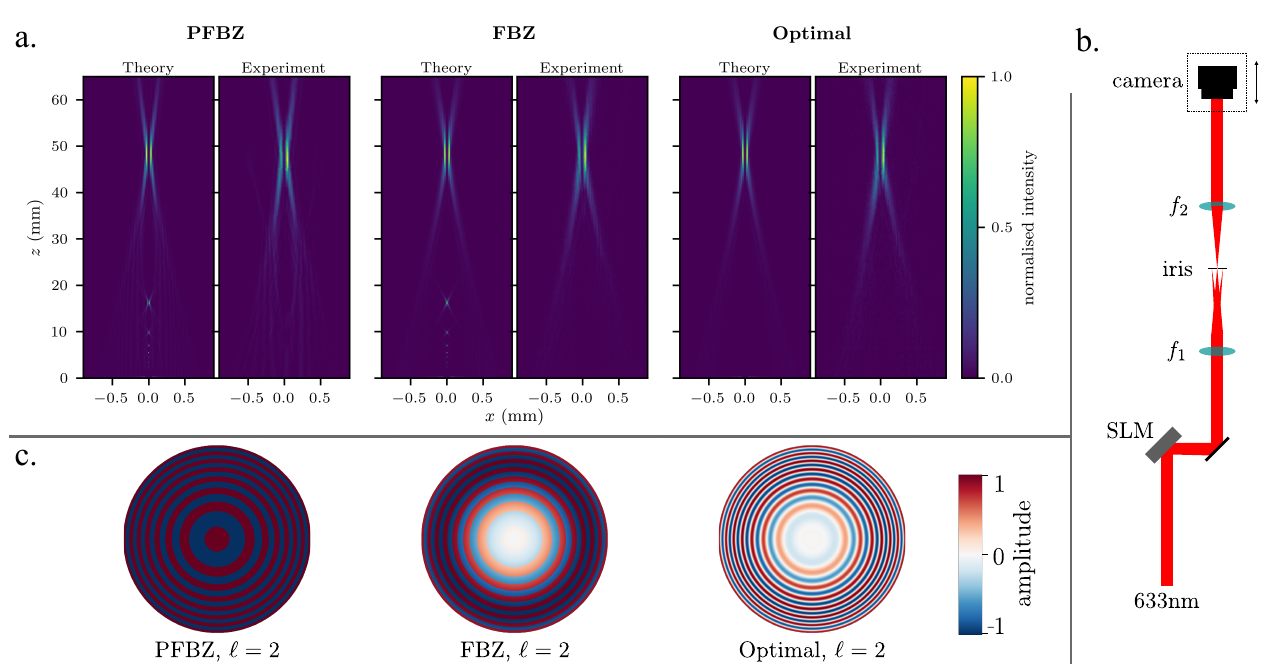}
    \caption{\textbf{a.} Theoretically calculated and experimentally measured transverse intensity distributions during free-space propagation for the corresponding masks, illustrating the formation and evolution of the vortex-ring focus for the case of $\ell=2$. \textbf{b.} Schematic of the experimental setup used to generate and analyze the optimal diffractive pupil functions. A linearly polarized Gaussian beam is expanded and modulated by a phase-only spatial light modulator (SLM), which encodes the calculated amplitude kinoforms using an amplitude--phase holographic technique. A $4f$ filtering system selects the desired diffraction order and removes unwanted components, producing the modulated vortex beam. The resulting field propagates in free space and is recorded with a CCD camera at different axial positions using a precise computer-controlled micro-stage. \textbf{c.} Representative examples of the calculated pupil masks used to generate ring focusing for topological charge $\ell=2$. The masks implement the optimal radial structure derived in the theory section. }
    \label{fig:setup}
\end{figure*}

By construction, $0\leq \eta(\psi)\leq 1$, and $\eta=1$ is attained only by the optimal pupil $\psi_\ell^{+}$ up to trivial degeneracies. 

To make the structure of the efficiency transparent, we decompose an arbitrary normalized pupil as $\psi=a_{\ell,+}\psi_{\ell}^{+}+a_{\ell,-}\psi_{\ell}^{-}+a_{\perp}\psi_\perp$, where $a_{\ell,\pm}=\langle\psi,\psi_{\ell}^{\pm}\rangle$, and $\langle \psi_\perp,\psi_{\ell}^{\pm}\rangle=0$. Since the kernel has support only on the active subspace $\mathrm{span}\{\psi_\ell^{+},\psi_\ell^{-}\}$, one has $\mathcal K_{\ell}\psi_\perp=0$. Substituting the decomposition into Eq.~\eqref{etadef}, immediately gives
\begin{equation}\label{eta2}
	\eta_{\ell}(\psi)=a_{\ell,+}^{2}+\frac{\Lambda^{(\ell)}_{-}}{\Lambda^{(\ell)}_{+}}\,a_{\ell,-}^{2}.
\end{equation}
\\
 
\noindent\textit{Fresnel--Bessel zones.} In the following, we apply Eq.(\ref{eta2}) to compare the performance of optimal solution $\psi^{+}_\ell(\rho)$ with that of Fresnel zone plate, consisting of $n$ zones for which $\tau_n=1/(2\pi n)$. For $\ell=0$, a direct calculation gives $a_{0,+}={2\sqrt{2}}/{\pi}$, $a_{0,-}=0$, $\Lambda^{(0)}_{+}=\Lambda^{(0)}_{-}=(\pi/2) n^2$, so that $\eta(\psi_{\mathrm{PFZ}})={8}/{\pi^{2}}$, which is in exact agreement with the results shown in Ref.~\cite{efremov2025optimal}. For nonzero OAM, $\ell\neq 0$, the notion of a Fresnel zone plate is no longer unique. The optimal pupil $\psi_\ell^{+}(\rho)$ contains both the Fresnel oscillation, $\cos\!\left[\rho^{2}/(2\tau_{\rm f})-\theta_\ell\right]$, and the OAM-dependent radial apodization, $J_\ell(\beta\rho)$ with $\beta={\rho^{*}}/{\tau_{\rm f}}$. This suggests two natural generalisations of the conventional binary Fresnel construction.

\paragraph*{(i) Binary Fresnel--Bessel zone mask.}
A direct extension is the binary pupil
\begin{equation}
	\psi_{\mathrm{PFBZ}}=2\sqrt{\pi}\,\mathrm{sign}\left[J_\ell(\beta\rho)\cos\left(\frac{\rho^{2}}{2\tau_{\rm f}}-\theta_{\ell}\right)\right].
\end{equation}
Its efficiency depends qualitatively on whether the Bessel factor changes sign within the aperture. A useful reference scale is the first zero $x_{\ell,1}$ of $J_\ell$. When $\beta<x_{\ell,1}$, the Bessel function does not change sign for $0\leq \rho\leq 1$. In the small-$\beta$ regime, one may use $J_\ell(\beta\rho)\approx (1/\ell!)\left(\beta\rho/2\right)^{\ell}$~\cite{arfken2011mathematical}. At the discrete focusing times $\tau_n=1/(2\pi n)$ and in the fast-Fresnel regime where the oscillatory phase varies rapidly compared with the Bessel envelope, averaging over the Fresnel oscillations gives
\begin{equation}\label{etabfz1}
	\eta_{\ell}(\psi_\mathrm{PFBZ})\approx\frac{8}{\pi^{2}}\frac{4(\ell+1)}{(\ell+2)^{2}}, \quad \beta\ll1\,.
\end{equation}
For $\ell=0$, Eq.~\eqref{etabfz1} reduces to $8/\pi^{2}$ \cite{efremov2025optimal}, while for $\ell>0$ the additional factor $4(\ell+1)/(\ell+2)^{2}<1$ quantifies the penalty associated with replacing the optimal Bessel apodization by a binary sign profile. This reduction is purely OAM-induced and has no analogue in the zero-charge case.

The behavior changes once $\beta>x_{\ell,1}$, where the Bessel factor itself develops sign reversals within the aperture. In the large-$\beta$ regime, the asymptotic form $J_\ell(\beta\rho)\approx\sqrt{2/(\pi\beta\rho)}\cos\left(\beta\rho-\ell\pi/2-\pi/4\right)$~\cite{arfken2011mathematical} shows that the pupil becomes the sign of a product of two rapidly oscillating factors, one Fresnel-like and one Bessel-like. In this limit, the efficiency approaches
\begin{equation}
	\eta(\psi_\mathrm{PFBZ})\approx\frac{512}{9\pi^{4}}, \quad \beta\gg1\,.
\end{equation}
Thus, once the Bessel oscillations are also binarized, the performance is further reduced relative to the small-$\beta$ regime.

\paragraph*{(ii) Amplitude-weighted Fresnel--Bessel mask.}
A second, less destructive generalization is to preserve the Bessel apodization while binarizing only the Fresnel oscillation. We therefore define
\begin{equation}
	\psi_{\mathrm{FBZ}}=\frac{J_\ell(\beta\rho)}{\sqrt{\mathcal I_0(\beta)}}\,\mathrm{sign}\left[\cos\left(\frac{\rho^{2}}{2\tau_{\rm f}}-\theta_{\ell}\right)\right]. 
\end{equation}
This construction retains the radial OAM weighting and therefore isolates the loss associated solely with Fresnel binarization. Defining $\chi(\rho)=\rho^{2}/(2\tau_{\rm f})-\theta_{\ell}$, the projections onto the optimal eigenmodes read $a_{\ell,\pm}^{2}=2A_{\ell,\pm}^{2}/\left[{\mathcal I_{\ell}(\mathcal I_{\ell}\pm|\mathcal R_{\ell}|)}\right]$ with 
\begin{align}\label{A_pm}
    A_{\ell,+}=&\frac{1}{2\pi}\int_0^1 \rho \dd\rho\,J_\ell^{2}(\beta\rho)|\cos\chi(\rho)|\\
    A_{\ell,-}=&\frac{1}{2\pi}\int_0^1 \rho \dd\rho\,J_\ell^{2}(\beta\rho)\, \mathrm{sign}[\cos\chi(\rho)]\sin\chi(\rho).
\end{align}
Substituting these relations into Eq.~\eqref{eta2} gives
\begin{equation}\label{etafbz}
	\eta_{\ell}(\psi_{\mathrm{FBZ}})=\frac{2\left(A_{\ell,+}^{2}+A_{\ell,-}^{2}\right)}{\mathcal I_{\ell}(\mathcal I_{\ell}+|\mathcal R_{\ell}|)}.
\end{equation}

A particularly transparent limit is the fast-Fresnel regime, in which $J_\ell^{2}(\beta\rho)$ varies slowly over one Fresnel period of $\chi(\rho)$. In this limit, $|\mathcal R_\ell|\approx 0$, $A_{\ell,+}\approx 2I_0/\pi$ and $A_{\ell,-}\approx 0$, so Eq.~\eqref{etafbz} reduces to $\eta(\psi_{\mathrm{FBZ}})\approx 8/\pi^{2}$. Thus, once the correct Bessel apodization is retained, the efficiency penalty associated with OAM disappears in the fast-Fresnel limit, and the remaining loss is exactly the familiar binary-harmonic penalty already encountered in the $\ell=0$ case.\newline

\begin{figure}
    \centering
    \includegraphics[width=\linewidth]{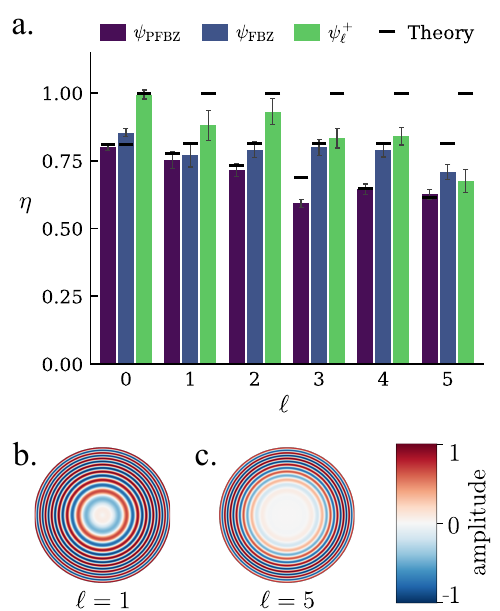}
    \caption{\textbf{a.} Experimental peak focal ring intensity for each OAM order $\ell=\{0,1,2,3,4,5\}$ for each pupil mask. Each intensity measurement is taken at the focal point of the mask, and averaged azimuthally about the highest intensity ring. This intensity is normalized by the relative input intensities for the different masks. We attribute the drop in efficiency of the optimal masks $\psi_{\ell}^{+}$ to the limited resolution of the phase-only SLM used in the experiment. As $\ell$ increases, the power is distributed more towards the higher frequency outer bounds of the mask, where the resolution of the SLM is less efficient. This can be seen in the optimal mask intensities shown, \textbf{b.} $\ell=1$, and \textbf{c.} $\ell=5$}
    \label{fig:exp-results}
\end{figure}

\noindent{\textit{Experimental implementation and results ---} To test the theory, we implemented optimal diffractive focusing of vortex beams in the optical domain. The experiment was designed to compare the propagation and ring-focusing performance of different pupil functions, including the binary Fresnel--Bessel mask $\psi_{\mathrm{PFBZ}}$ and the amplitude-weighted Fresnel--Bessel mask $\psi_{\mathrm{FBZ}}$, with the corresponding optimal target profile. The simplified version of the sketch experimental setup is shown in Fig.~\ref{fig:setup}. A linearly polarized Gaussian beam was first expanded to match the active area of a reflective liquid-crystal-on-silicon spatial light modulator (Hamamatsu, $800\times600$ pixels, pixel pitch $12.5~\mu$m). The spatial light modulator (SLM) was used to encode the desired real-valued pupil functions by means of an amplitude-phase modulation scheme~\cite{bolduc2013exact}. In the case of the binary mask $\psi_{\mathrm{PFBZ}}$, the hologram directly encoded the sign-changing Fresnel--Bessel structure. For the amplitude-weighted mask $\psi_{\mathrm{FBZ}}$, the hologram additionally incorporated the required radial apodization associated with the Bessel envelope. In both cases, the calculated pupil functions were converted into phase-only holograms using an amplitude-phase encoding protocol, allowing the target real modulation to be synthesized with the phase-only SLM. After modulation, the first diffraction order was spatially selected using an iris at the Fourier plane of a $4f$ filtering system, thereby removing the unwanted zero- and higher-order diffraction orders generated by the holographic encoding. The selected beam was then reimaged and allowed to propagate, and its transverse intensity distribution was recorded with a CCD camera at different axial positions using a precise, computer-controlled micro-stage. This procedure enabled direct measurement of the propagation dynamics, in particular, the formation of the target annular focus. The experimentally recorded intensity patterns were then compared with the theoretical propagation profiles obtained from Eq.~\eqref{psiprop}. Representative transverse intensity distributions at several propagation planes are shown in  Fig.~\ref{fig:setup}, together with the corresponding calculated profiles. In all cases, the observed vortex-ring evolution closely follows theoretical expectations, confirming that the encoded holograms reproduce the intended diffractive pupils with high fidelity. To quantify the focusing performance, we extracted the peak intensity at the target ring as a function of propagation distance and compared the results for the two mask families, shown in Fig.~\ref{fig:exp-results}. The measurements demonstrate that the amplitude-weighted Fresnel--Bessel mask $\psi_{\mathrm{FBZ}}$ yields a systematically higher ring intensity than the fully binary Fresnel--Bessel mask $\psi_{\mathrm{PFBZ}}$, in agreement with the theoretical efficiency analysis presented above. This difference is consistent with the role of the Bessel apodization factor: when the radial OAM-dependent envelope is retained, the remaining loss is dominated by the binary Fresnel modulation alone, whereas full binarization introduces an additional penalty associated with the improper radial weighting.
Overall, the experimental results validate the central theoretical prediction that optimal diffractive focusing of OAM beams is governed not only by Fresnel phase matching, but also by the Bessel-type radial structure imposed by the vortex geometry. They further show that holographically implemented pupil functions provide a direct and flexible route for testing the fundamental limits of ring focusing in structured beams.}

\textit{Conclusion---} In this work, we have extended the optimal aperture-limited focusing theory of Ref.~\cite{efremov2025optimal} to fields carrying non-zero orbital angular momentum. Unlike conventional focusing, vortex beams possess a phase singularity that imposes vanishing on-axis intensity, so the natural focusing objective is to concentrate energy onto a ring rather than a point. As a result, the optimal pupil is no longer determined solely by Fresnel phase matching, which is responsible for the target focal point, but also exhibits a radial Bessel structure dictated by the OAM charge and the target ring radius. Despite this additional structure, the optimization problem remains analytically tractable. We showed that the focusing operator retains an effectively two-dimensional active subspace, which allows the optimal pupil, the maximum achievable ring intensity, and the performance of arbitrary pupil functions to be determined in closed form. This structure provides a natural benchmark for evaluating practical diffractive masks and clarifies how deviations from the optimal pupil degrade the achievable focusing performance. A key outcome of this analysis is that, for $\ell\neq0$, the limitations of conventional Fresnel-type masks arise from two distinct sources. In addition to the well-known efficiency loss associated with binary-phase quantization, vortex-beam focusing also requires a specific radial apodization determined by the Bessel envelope. Masks that fail to reproduce this OAM-dependent radial weighting suffer an additional penalty that has no analogue in the $\ell=0$ case. By contrast, designs that preserve the Bessel modulation while approximating the Fresnel oscillation recover the universal binary-harmonic efficiency limit. The experimental implementation using spatial light modulation confirms these theoretical predictions and demonstrates that optimal diffractive pupils can be realized and directly compared with practical mask families. More broadly, our results establish the fundamental limits of diffractive focusing for vortex beams and provide a systematic framework for designing high-efficiency ring-focusing elements for structured waves. These ideas are not restricted to optical beams. Because the formulation is expressed entirely in terms of aperture-limited propagation, the same framework applies to a wide class of wave systems, including electron and neutron beams as well as other structured matter waves. The results, therefore, provide a general route for analyzing and engineering optimal diffractive focusing in systems where conventional refractive optics is unavailable or impractical.\\

\noindent \textbf{Authors' contributions.} R.G. conceived the idea and developed the theory through discussion with E.K and M.A.E. R.G performed the simulations. C.R. and L.S. conducted the experiments and analyzed the data. M.A.E. and E.K. supervised the project. R.G and E.K. prepared the first draft. All authors contributed to the manuscript writing.\\
\noindent \textbf{Acknowledgments.}
E.K. acknowledges the support of Canada Research Chairs; National Research Council of Canada Quantum Sensing Program.\\
\noindent \textbf{Disclosures} The authors declare no conflicts of interest.\\
\noindent \textbf{Data availability} Data from the results presented in this paper may be obtained from the authors upon reasonable request.

\bibliographystyle{apsrev4-1}
\bibliography{ref.bib}

\appendix
\section{Additional derivations}

\setcounter{equation}{0}
\renewcommand{\theequation}{A.\arabic{equation}}

\textbf{Derivation of the OAM-projected Fresnel propagator.} We start from the dimensionless free Schr\"odinger (paraxial) equation
\begin{equation}
\label{Schr-equation}
\ii\frac{\partial}{\partial \tau} \Psi(\boldsymbol{\rho},\tau)= -\frac{1}{2}\nabla_{\boldsymbol{\rho}}^2\,\Psi(\boldsymbol{\rho},\tau)
\end{equation}
for the function $\Psi(\boldsymbol{\rho},\tau)$, where $\boldsymbol{\rho}=(\rho, \varphi)$, with the initial condition $\Psi(\boldsymbol{\rho},0)=\Psi_0(\boldsymbol{\rho})$. Introducing the two-dimensional Fourier transform with respect to spatial coordinate $\boldsymbol{\rho}$
$$
\widetilde{\Psi}(\mathbf{q},\tau)=\int_{\mathbb{R}^2} \dd^2\boldsymbol{\rho}\;\ee^{-\ii\mathbf{q}\cdot\boldsymbol{\rho}}\Psi(\boldsymbol{\rho},\tau)
$$
the partial differential equation \eqref{Schr-equation} becomes an ordinary differential equation
$$
\ii\frac{\partial}{\partial \tau} \widetilde{\Psi}(\mathbf{q},\tau)=\frac{q^2}{2}\,\widetilde{\Psi}(\mathbf{q},\tau)
$$
with the solution $\widetilde{\Psi}(\mathbf{q},\tau)=\ee^{-\ii q^2\tau/2}\,\widetilde{\Psi}_0(\mathbf{q})$. Here we have used the initial condition for the function $\Psi(\boldsymbol{\rho},\tau)$.

The inverse Fourier transformation determines the solution
\begin{equation}
\Psi(\boldsymbol{\rho},\tau)=\int_{\mathbb{R}^2} \dd^2\boldsymbol{\rho}'\;G(\boldsymbol{\rho}-\boldsymbol{\rho}',\tau)\,\Psi_0(\boldsymbol{\rho}'),
\end{equation}
of Eq. \eqref{Schr-equation} with the initial condition in terms of the Green's function
\begin{equation}
G(\boldsymbol{\rho},\tau)=\frac{1}{(2\pi)^2}\int_{\mathbb{R}^2} \dd^2\mathbf{q}\,\ee^{\ii\mathbf{q}\cdot \boldsymbol{\rho}-\frac{\ii}{2}q^2\tau}
=\frac{1}{2\pi \ii\tau}\exp\!\left(\frac{\ii\boldsymbol{\rho}^2}{2\tau}\right),
\end{equation}
giving rise to 
\begin{equation}
\Psi(\boldsymbol{\rho},\tau)=\frac{1}{2\pi \ii\tau}\int_{\mathbb{R}^2} \dd^2\boldsymbol{\rho}'\,
\exp\!\left(\frac{\ii\,|\boldsymbol{\rho}-\boldsymbol{\rho}'|^2}{2\tau}\right)\,\Psi_0(\boldsymbol{\rho}').
\end{equation}

Next we assume the initial state carries a single OAM charge $\Psi_0(\rho',\varphi')=\psi_\ell(\rho')\,\ee^{\ii\ell\varphi'}$ with $\psi_\ell(\rho')\in\mathbb{R}$. Substituting this form into the propagator, restricting the initial field to unit pupil, $\rho'\le 1$, and using the relation $|\boldsymbol{\rho}-\boldsymbol{\rho}'|^2=\rho^2+\rho'^2-2\rho\rho'\cos(\varphi-\varphi')$, with $\boldsymbol{\rho}'=(\rho',\varphi')$, we arrive at
\begin{eqnarray}
\Psi(\rho,\varphi,\tau)&=&\frac{\ee^{\frac{\ii}{2\tau}\rho^2}}{2\pi \ii\tau}\int_0^1 \rho' \dd\rho'\;\ee^{\frac{\ii}{2\tau}\rho'^2}\psi_\ell(\rho')\cr
&\times&\int_0^{2\pi} \dd\varphi'\,\exp\!\left[-\ii\,\frac{\rho\rho'}{\tau}\cos(\varphi-\varphi')\right]e^{\ii\ell\varphi'}.
\end{eqnarray}

Applying the Jacobi-Anger expansion $\ee^{\ii x\cos\theta}=\sum_{n=-\infty}^{\infty}\ii^n J_n(x)\,\ee^{\ii n\theta}$, the angular integral yields
\begin{align}
\int_0^{2\pi} \dd\varphi'\,
\exp\!\left[-\ii\,\frac{\rho\rho'}{\tau}\cos(\varphi-\varphi')\right]\,\ee^{\ii\ell\varphi'}
&=2\pi\,(-\ii)^\ell\,\ee^{\ii\ell\varphi}\,J_\ell\left(\frac{\rho\rho'}{\tau}\right),
\end{align}
where we used the definition $\int_0^{2\pi} \dd\varphi'\,\ee^{\ii(\ell-\ell')\varphi'}=2\pi\,\delta_{\ell\ell'}$ of the Kronecker symbol $\delta_{\ell\ell'}$. As a result, we obtain $\Psi(\rho,\varphi,\tau)=\ee^{\ii\ell\varphi}\Psi(\rho,\tau)$, with $\Psi(\rho,\tau)$ given by the Fresnel-Bessel propagation integral 
\begin{equation}\label{psiprol}
\Psi(\rho,\tau)
=\frac{(-\ii)^\ell}{\ii\tau}\,\ee^{\ii\frac{\rho^2}{2\tau}}
\int_0^1 \rho' \dd\rho'\,\ee^{\ii\frac{\rho'^2}{2\tau}}\,
J_\ell\left(\frac{\rho\rho'}{\tau}\right)\,\psi_\ell(\rho').
\end{equation}
which is Eq.(\ref{psiprop}).

\textbf{Solution of Eq.(\ref{eigenvalue}).} Using $\cos(a-b)=\cos a\cos b+\sin a\sin b$, Eq.(\ref{kernel}) separates into
\begin{equation}
	\mathcal K_{\ell}(\rho,\rho')=\frac{1}{\tau_{\rm f}^2}\Big[f_c(\rho)f_c(\rho')+f_s(\rho)f_s(\rho')\Big]
\label{kernel2},
\end{equation}
with
\begin{eqnarray}\label{fcs}
	f_c(\rho)&=&J_\ell\!\left(\frac{\rho^* \rho}{\tau_{\rm f}}\right)\cos\!\left(\frac{\rho^2}{2\tau_{\rm f}}\right),\cr\nonumber
	f_s(\rho)&=&J_\ell\!\left(\frac{\rho^* \rho}{\tau_{\rm f}}\right)\sin\!\left(\frac{\rho^2}{2\tau_{\rm f}}\right).
\end{eqnarray}

Introducing the inner product $\langle g,h\rangle=\frac{1}{2\pi}\int_0^{1}\rho\,\dd\rho\, g(\rho)h(\rho)$, Eq.~\eqref{eigenvalue}, becomes
\begin{equation}\label{eigenproblem}
	\frac{1}{\tau_{\rm f}^2}\left[f_c(\rho)\langle f_c,\psi_\ell\rangle+f_s(\rho)\langle f_s,\psi_\ell\rangle\right]=\Lambda\,\psi_\ell(\rho).
\end{equation}
The kernel, therefore, maps any pupil function into the two-dimensional subspace spanned by $\{f_c,f_s\}$ given in Eq.~\eqref{fcs}. Consequently, the eigenfunctions can be sought within this subspace, as $\psi_\ell(\rho)=u f_c(\rho)+v f_s(\rho)$. Defining the overlaps
$A=\langle f_c,f_c\rangle$, $B=\langle f_s,f_s\rangle$, and $C=\langle f_c,f_s\rangle$, the eigenvalue problem reduces to the $2\times2$ matrix equation
\begin{equation}\label{eigenproblem2}
	\frac{1}{\tau_{\rm f}^2}
	\begin{pmatrix}
	A & C\\
	C & B
	\end{pmatrix}
	\begin{pmatrix}
	u\\
	v
	\end{pmatrix}
	=
	\Lambda\,
	\begin{pmatrix}
	u\\
	v
	\end{pmatrix}.
\end{equation}

Solving Eq.~\eqref{eigenproblem2} gives the eigenvalues and corresponding eigenfunctions appearing in Equations~\eqref{jmax}-\eqref{psipml}.

\begin{figure*}
    \centering
    \includegraphics[width=0.9\textwidth]{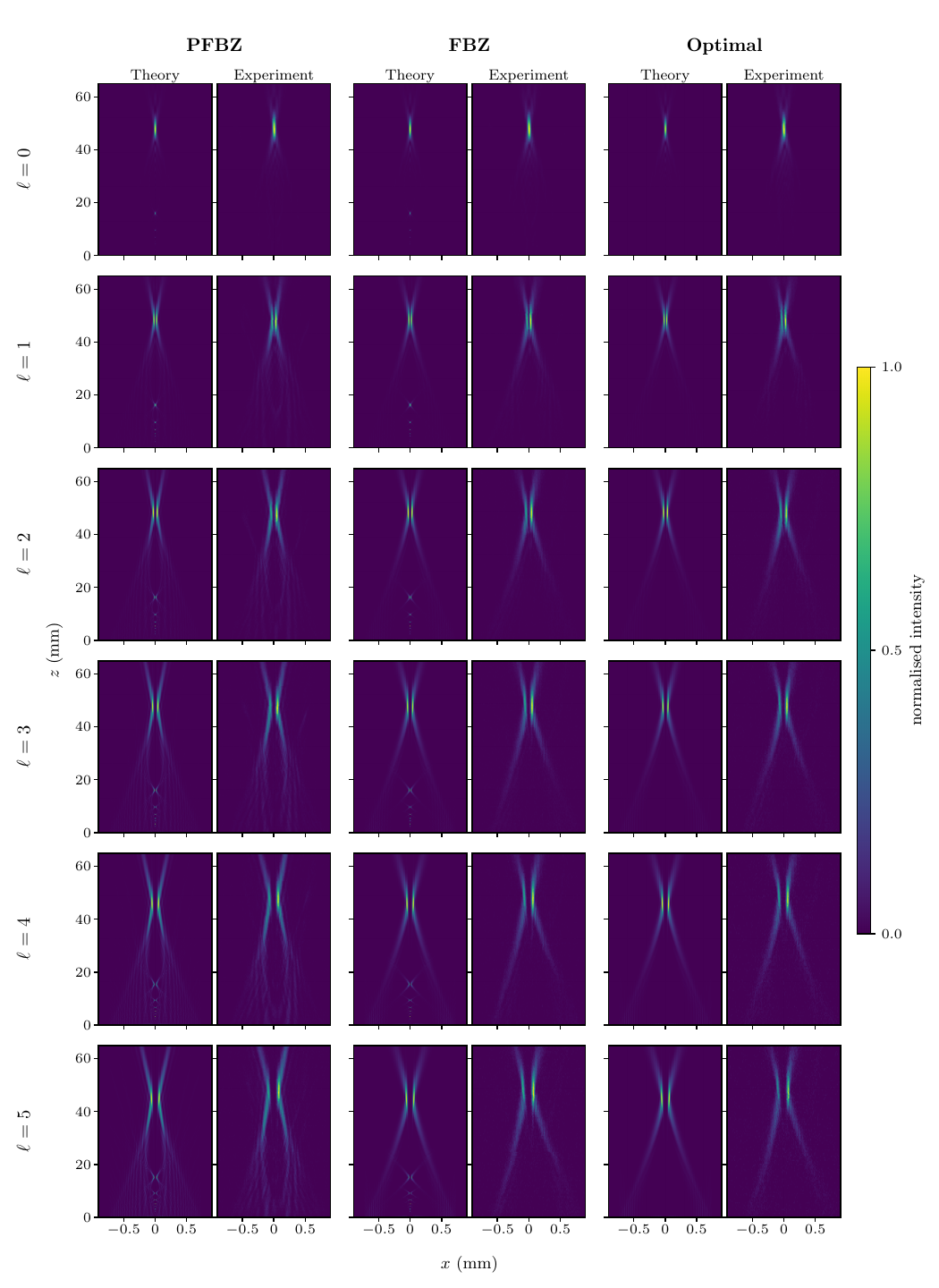}
    \caption{Theoretically calculated and experimentally measured transverse intensity distributions during free-space propagation for all masks and beams possessing OAM or 0--5.}
    \label{fig:Supp1}
\end{figure*}

\begin{figure*}
    \centering
    \includegraphics[width=0.9\textwidth]{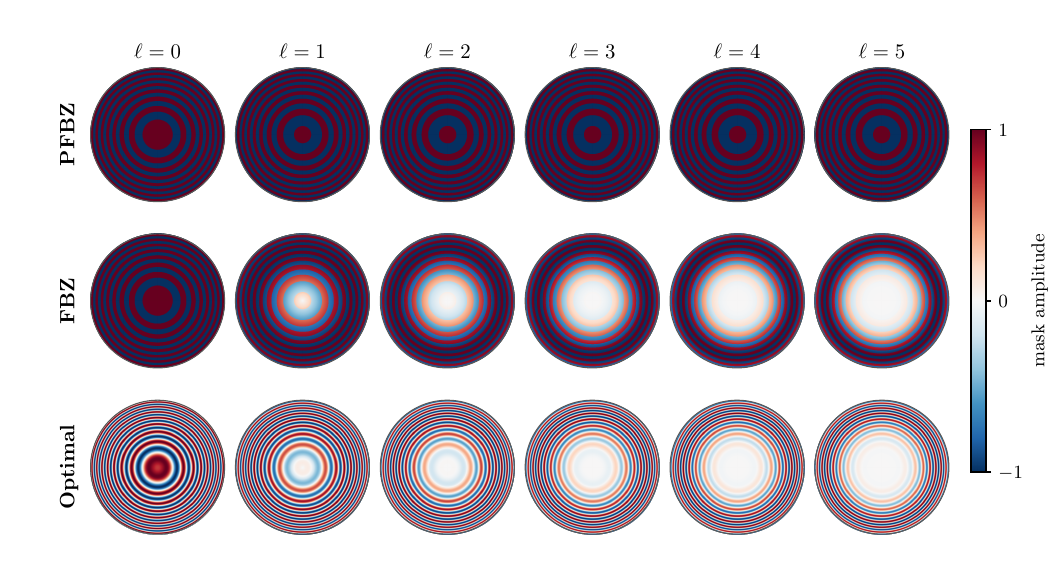}
    \caption{Mask amplitudes for all masks, normalized so that the maximum value is one.}
    \label{fig:Supp2}
\end{figure*}

\end{document}